\documentclass[sigconf]{acmart}

\AtBeginDocument{%
  \providecommand\BibTeX{{%
    \normalfont B\kern-0.5em{\scshape i\kern-0.25em b}\kern-0.8em\TeX}}}

\setcopyright{acmcopyright}
\copyrightyear{2021}
\acmYear{2021}
\acmDOI{10.1145/1122445.1122456}

\usepackage[ruled,linesnumbered]{algorithm2e}

\begin{document}

\title{Multiple Interest and Fine Granularity Network for User Modeling}

\author{Jiaxuan Xie, Jianxiong Wei, Qingsong Hua, Yu Zhang}

\email{{xiejiaxuan.xjx, jianxiong.wjx}@alibaba-inc.com, {qingsong.huaqs, daoji}@taobao.com}

\affiliation{%
  \institution{Alibaba Group}
  \city{Beijing}
   \country{China}
}








 \renewcommand{\shortauthors}{Jiaxuan Xie, et al.}

\begin{abstract}
User modeling plays a fundamental role in industrial recommender systems, either in the matching stage and the ranking stage, in terms of both the customer experience and business revenue. How to extract users' multiple interests effectively from their historical behavior sequences to improve the relevance and personalization of the recommend results remains an open problem for user modeling. 
Most existing deep-learning based approaches exploit item-ids and category-ids but neglect fine-grained features like color and material, which hinders modeling the fine granularity of users' interests. 
In the paper, we present Multiple interest and Fine granularity Network (MFN), which  tackle users' multiple and fine-grained interests and construct the model 
from both the similarity relationship and the combination relationship among the users' multiple interests.
Specifically, for modeling the similarity relationship, we leverage two sets of embeddings, where one is the fixed embedding from pretrained models (\textit{e.g.} Glove) to give the attention weights and the other is trainable embedding to be trained with MFN together. For modeling the combination relationship, self-attentive layers are exploited to build the higher order combinations of different interest representations. In the construction of network, we design an interest-extract module using attention mechanism to capture multiple interest representations from user historical behavior sequences and leverage an auxiliary loss to boost the distinction of the interest representations. Then a hierarchical network is applied to model the attention relation between the multiple interest vectors of different granularities and the target item. We evaluate MFN on both public and industrial datasets. The experimental results demonstrate that the proposed MFN achieves superior performance than other existed representing methods.

\end{abstract}




\maketitle

\section{Introduction}
In recent years, recommender systems have become increasingly prevalent and gained success in various applications such as news distribution, video watching and e-commerce. Instead of only considering categories or keywords, current recommender approaches attempt to merge more personalized information into modeling with the goal of understanding exactly the intention of users and showing what they are most interested in. The user historical behavior
sequences, which have been proved to be of great value for personalization, play a significant role in user modeling for extracting users' hidden interests.

A popular user modeling strategy is to obtain user interest representations by leveraging the mean pooling or the weighted pooling over the historical behavior sequences. Several methods follow a similar Embedding\&MLP paradigm \cite{DBLP:conf/recsys/CovingtonAS16, Cheng0HSCAACCIA16, ZhaiCZZ16}. 
They capture the representation of users' interests by averaging the embedding vectors of user behaviors and transform them into a fixed-length vector as users' interest representation. While attention-based methods like
DIN \cite{ZhouZSFZMYJLG18}, DIEN \cite{ZhouMFPBZZG19} and DSIN \cite{FengLSWSZY19}, they adaptively  extract the insterest vector by considering the relevance of historical behaviors given a target item using attention mechanisms, which consequently assigns higher activated weights to those historical behaviors with higher relevance. DIN and its successors achieve better performance than Embedding\&MLP methods, but a single interest vector is still insufficient to capture the varying characteristics of users' interests. To address this problem, MIND \cite{DBLP:conf/cikm/LiLWXZHKCLL19} and ComiRec \cite{DBLP:conf/kdd/CenZZZYT20}  leverage capsule routing mechanism for clustering historical behaviors and obtaining users' multiple interest vectors.
However, except item-ids, category-ids and similar id features, both of them lack the modeling of finer grained features like color and materials. \cite{DBLP:conf/cikm/LvJYSLYN19,DBLP:conf/cikm/GuDWZLY20, DBLP:conf/cikm/XiaoYJWHW20} exploit transformers or similar hierarchical attention structure to capture multiple interest representations of users. Yet, their complicated model structures make serving online nearly infeasible without specific optimization for practical usage. 

In this paper, we propose Multiple Interest and Fine granularity  Network (MFN), for user modeling to handle with users' multiple and fine-grained interests. There are two key components in MFN, one is for modeling the similarity relationship and the other is for the combination relationship among users' multiple interests. \textbf{Similarity} means two interests are similar in terms of physical characteristics, $\textit{e.g.}$ a customer likes black windbreakers and black coats. \textbf{Combination} means two interests share latent collocation,  $\textit{e.g.}$ a father buys beers and pampers. Specifically, for modeling the similarity relationship, we leverage two sets of embeddings, where one is the fixed embedding from pretrained models (\textit{e.g.} Glove) to give the attention weights and the other is trainable embedding to be trained with MFN together. For modeling the combination relationship, self-attentive layers are exploited to build the higher order combinations of different interest representations. By modeling the similarity and the combination relationship separately,  we are able to build more expressive and effective users' multiple and fine-grained interest representations. In a nutshell, the main contributions of this paper can be concluded as following:
\begin{itemize}
    \item We propose to study the problem of extracting multiple and fine-grained interest representations from users' historical behavior sequences for user modeling and present the Multiple interest and Fine granularity Network (MFN).
    \item We propose to model both the similarity relationship and the combination relationship among users' multiple interests. For the former, we leverage two sets of embeddings, where the fixed one is for giving he attention weights and the trainable one is trained for the main task. For the latter, we leverage self-attentive layers to learning high-order interest interactions.
    
    \item We conduct extensive experiments on both public and industrial datasets. Experimental results demonstrates the proposed MFN achieves superior performance than other representing methods with other important attributes like good explainability and little online response time.
\end{itemize}

\section{Related Work}
In recent decades, user interest modeling has attracted much attention in industrial applications such as recommender systems and online advertising, which concentrates on learning the representation of users' interests from the historical behavior sequences. DIN \cite{ZhouZSFZMYJLG18} leverages an attention mechanism to capture the diverse interests of a user on different candidate items.
DIEN \cite{ZhouMFPBZZG19} considers the temporal relationship among  the historical behaviors and proposes to model the evolution of users' interests with an interest extraction layers based on GRU. DSIN \cite{FengLSWSZY19} highlights that user behaviors are highly homogeneous in each session and heterogeneous cross sessions and designs a self-attention network with bias encoding to get the corresponding interest representation of each session. MIND \cite{DBLP:conf/cikm/LiLWXZHKCLL19} 
and ComiRec \cite{DBLP:conf/kdd/CenZZZYT20} emphasize that a single interest vector is insufficient to capture the varying nature of users' interests. They exploit capsule network and dynamic routing to obtain the representation of users' interests as multiple interest vectors. Due to the success of transformers in other areas, \cite{DBLP:conf/cikm/SunLWPLOJ19, DBLP:conf/cikm/LvJYSLYN19} propose to leverage transformers or similar structures with multiple head attentions to extract users' multiple interests from the behavior sequences.

There are also a series of works that focus on long-term even lifelong user interest modeling. \cite{DBLP:conf/kdd/PiBZZG19} demonstrates that the longer historical behavior sequences in the user interest modeling module can support the CTR model with better performance. MIMN \cite{DBLP:conf/kdd/PiBZZG19} 
embeds user long-term interest into fixed-sized memory network to decrease the burden of the latency and storage of  online serving. SIM \cite{DBLP:conf/cikm/sim} leverages a general search unit to get a sub user behavior sequence that is relevant to  the candidate item and proposes an exact search unit to model the precise relationship between the candidate item and the sub sequence.

\section{methodology}
In this section, we formulate the problem and illustrate the proposed nt Multiple interest and Fine granularity Network (MFN) in detail. Considering the deployment of MFN in practical scenarios, we build MFN with a hierarchical structure with two levels. The first level is for category and the second level is for fine-grained features like item entities.

\subsection{Preliminary}
Suppose we have a set of users $\mathcal{U}$ and a set of items $ \mathcal{I}$. For simplicity, here we consider a user $u \in \mathcal{U}$ and a candidate item $i_c \in \mathcal{I}$. The historical behavior sequence of $u$ is $E = \left( i_1, \ldots, i_t,\ldots, i_N     \right) \in \mathbb{R}^{N\times d}$, where  $E  \subset \mathcal{I}$, $t$ denotes the $t$-th interaction item $d$ is the feature size,  $N$ is the length of the sequence.  The feature fields of items include id features ( \textit{e.g.} iid-item id, cid-category id, sid-shopid, bid-brand id) and fine-grained features (\textit{e.g.} entity features like color or materials), \textit{i.e.} $ i_t =\left(i_{t}^{coarse-id};i_{t}^{fine} \right)$
, where $i_{t}^{coarse-id} \subseteq \{i_{t}^{iid}, i_{t}^{cid}, i_{t}^{sid}, i_{t}^{bid}\}, i_{t}^{fine}$ denote the features of coarse id features and fine-grained features, respectively. The feature construction here follows the basic embedding paradigm \cite{CovingtonAS16, Cheng0HSCAACCIA16}. 

\begin{figure}[h]
  \centering
  \includegraphics[width=\linewidth]{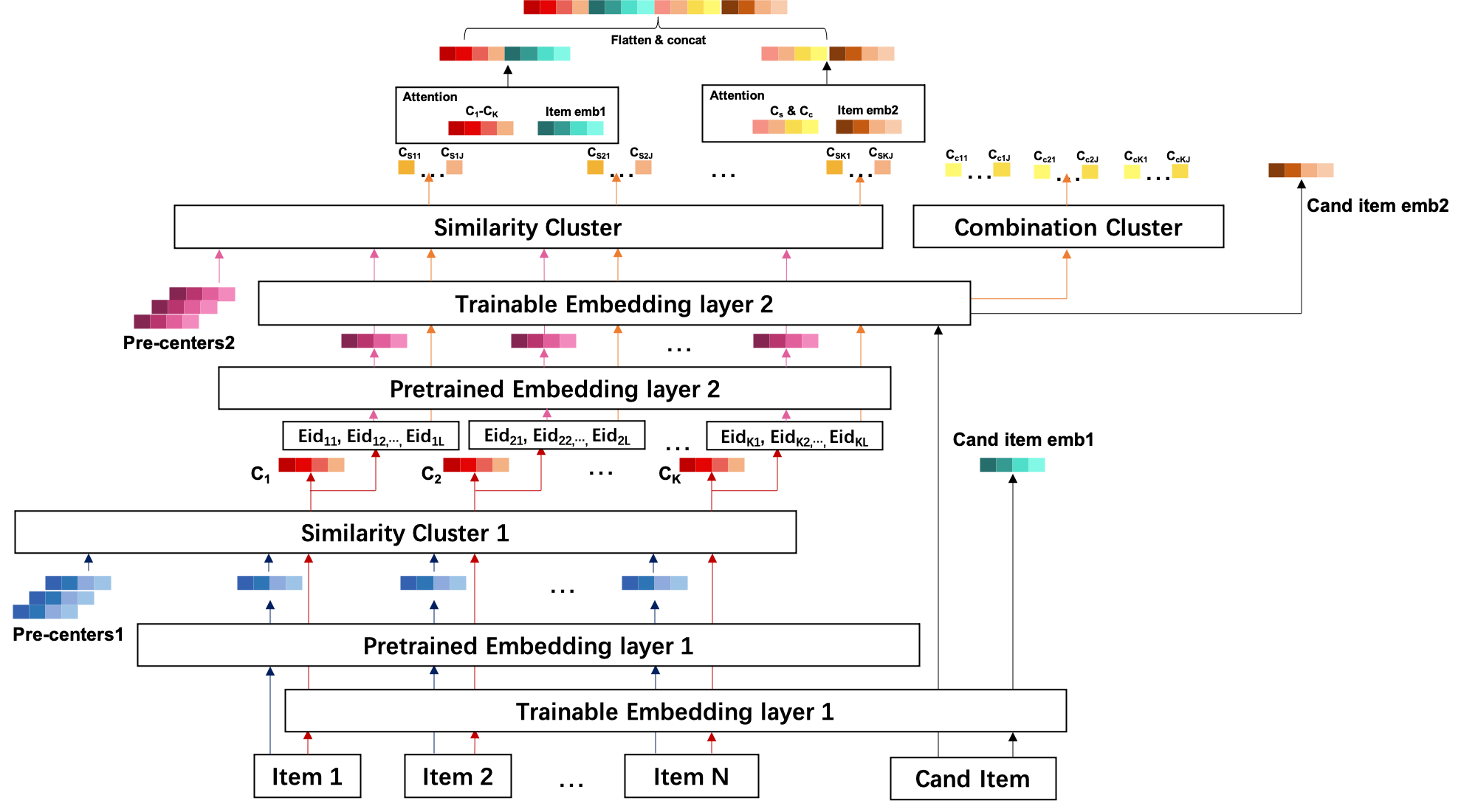}
  \caption{The structure of MFN.}

\end{figure}

\subsection{Similarity}
Similarity means two interests are similar in terms of physical characteristics, \textit{e.g.} a customer likes black windbreakers and black coats. Hence, we can divide and extract users' multiple interests according to the similarity of the items in the behavior sequences.

Here, to model the similarity relationship among the users' multiple interests, we construct two sets of embeddings, $E$ and $\mathcal{E}$, $E, \mathcal{E} \in \mathbb{R}^{N\times d}$. One is provided by pretrained models and is fixed in the model to measure the similarity, still denoted as $E$. For id features like iid, cid, sid, the pretrained embedding can be obtained from a vanilla model under an auxiliary CTR prediction task based on long-term behavior data. If the fine-grained features are from words, we can exploit Golve or Word2Vec to get the pretrained embeddings. And the other set of embedding $\mathcal{E}$ is trainable and is trained under the main task in the proposed MFN model. For clustering multiple interest vector centers,  \cite{DBLP:conf/cikm/LiLWXZHKCLL19, DBLP:conf/kdd/CenZZZYT20} leverage random centers as the initial centers. However, Using random centers not only does require more iteration in the following clustering, but also lose the expressive power and the explainability to some degree. In this paper, we propose to pretrain the cluster centers. Suppose one user's multiple interest cluster centers  are $C = \{ C_1, \ldots, C_j  \ldots, C_K \} \in \mathbb{R}^{K \times d}$, where $K$ is the number of centers, $C_j$ denotes the $j$-th center. Recall the historical behavior sequence $E = \left( i_1, \ldots, i_t,\ldots, i_N     \right) \in \mathbb{R}^{N\times d}$, we assume the probability that behavior $i_t$ belongs to center $C_j$ is $P_{tj}$. On the one hand, we hope that the centers are not redundant, $\textit{i.e.}$ each cluster is filled with samples and each  center can be covered, hence we maximum the entropy of the sum along the rows of $P$, 
\begin{equation}
    \max \mathcal{L}_{ME} := -\frac{1}{K}\sum_{j=1}^{K}\left( (\frac{1}{N}\sum_{t=1}^{N}P_{tj})\log (\frac{1}{N}\sum_{t=1}^{N}P_{tj})\right).
\end{equation}

On the other hand, one behavior $i_t$ ought to be assigned to only one centers, hence we minimum the entropy of each $P_{tj}$,
\begin{equation}
    \min \mathcal{L}_{SE} :=-\frac{1}{K}\frac{1}{N} \sum_{j=1}^{K}\sum_{t=1}^{N}\left(P_{tj}\log (P_{tj})\right).
\end{equation}

From Eqn. (1) and  Eqn. (2), the optimization target is to minimize
\begin{equation}
    \mathcal{L}_E = \mathcal{L}_{SE}- \mathcal{L}_{ME}.
\end{equation}

Considering the large-scale situation, we can use the mini-batch version.  The algorithm to obtain the multiple interest centers is demonstrated as Alg. \ref{algorithm1}. Note that here we leverage backward propagation to optimize the trainable centers $C$, and the stopping criterion can be set to stop at a given number of iterations.
\begin{algorithm}
\caption{The optimization of multiple interest centers}
\label{algorithm1}
\KwData{Trainable centers $C\in \mathbb{R}^{K \times d}$, $M$ users' historical behavior sequence $E^{all} = \{E^l\}_{l=1}^{M}, E^l \in \mathbb{R}^{N\times d} $, learning rate $\alpha$}
\KwResult{Trained centers $C$.}
Initialize $\mathcal{L}_E=0.0$\;
\While{convergence criterion not satisfied}
{
batch $B$ $\leftarrow$ uniform random sample of size $b$ from $\{1,\ldots,M \}$\;
\For{all $ k \in B$}
{
    $P^k = E^k{C}^{\top}$ with entries $P^k_{tj}=E^k_t(C_j)^{\top}$\;
    $P^k \leftarrow \textrm{softmax}(P^k)$\;
    Calculate $\mathcal{L}_E^k$ from Eqn. (1), (2) and (3)\;
    $\mathcal{L}_E\leftarrow \mathcal{L}_E+\mathcal{L}_E^k$\;
}
Compute the gradient $\delta C$ from the backward of $\mathcal{L}_E$\;
Update $C\leftarrow C-\alpha\delta C$\;
}
\end{algorithm}





And in the next we introduce how to leverage the multiple interest centers in the main task of MFN. After obtaining the multiple interest centers $C$, using attention mechanism  we are able to give the similarity weights, $i.e.$ the probabilities that the behaviors $\mathcal{E}^l$ can be divided into interest centers $C$,
\begin{equation}
    P^l = \textrm{softmax}(E^{l}C^{\top}).
\end{equation}
And the multiple interest representations of a user $l$ is 
\begin{equation}
    R_s^{l} = (P^l)^{\top}\mathcal{E}^l,
\end{equation}
where $R_s^{l} \in \mathbb{R}^{K \times d}$.

\subsection{Combination}
Besides the similarity attribute in terms of physical or semantic characteristics,
another significant relationship among the  users' multiple interests is combination. Combination here means two interests share latent collocation, \textit{e.g.} a father buys beers and pamper at the same time, or a student buys an English book and an electronic dictionary. Here, we adapt the self attentive method \cite{DBLP:conf/iclr/LinFSYXZB17} for capturing high-order combinations of multiple interests. Recall $\mathcal{E} \in \mathbb{R}^{N\times d}$ is the trainable embedding for behavior sequences, using self-attention mechanism, we get the combinational weight matrix $A$ as 
\begin{equation}
    A = \textrm{softmax}(\mathbf{W}_2^{\top}\textrm{swish}(\mathbf{W}_1 (\textrm{MSA}(\mathcal{E}))^{\top})),
\end{equation}
where $\textrm{swish}(\cdot)$ denotes the $
\textrm{swish}$ activation, $\textrm{MSA}(\cdot)$ denotes the Multi-head self-attention, $\mathbf{W}_1$ and $\mathbf{W}_2$ are  trainable parameters with size $d_h \times d$ and $d_h \times K$, respectively. 

The combinational interest representations $R_c$ can be computed by
\begin{equation}
    R_c = A\mathcal{E},
\end{equation}
where $R_c \in \mathbb{R}^{K \times d}$, $K$ is the number of interests and $d$ is the dimension of features.

\subsection{Aggregation}
In this section, we introduce how to aggregate the interest representations from both the similarity and combination relationships.

After extracting the multiple interests for each user from the historical behavior sequences, we have the whole interests $\mathcal{R}$ as
\begin{equation}
    \mathcal{R} = [R_s^l; R_c^l],
\end{equation}
where $\mathcal{R} \in \mathbb{R}^{2K \times d}$. 
We can get the interest representation $\mathcal{R}(i_c)$ with regard to the candidate item $i_c \in \mathbb{R}^{1\times d}$,
\begin{align}
    \mathcal{R}(i_c) = f(\mathcal{R}, i_c) \\
    &= \sum_{j=1}^{K} a_1(\mathbf{r}_j, i_c)\mathbf{r}+
    \sum_{j=K}^{2K} a_2(\mathbf{r}_j, i_c)\mathbf{r} \\
     &=  \sum_{j=1}^{K}w_{1j} \mathbf{r}_j+\sum_{j=K}^{2K}w_{2j} \mathbf{r}_j
\end{align}
where $\mathbf{r}_j$ denotes the $j$-th interest in $\mathcal{R}$ ($\textit{i.e.}$ the $j$-th row of $\mathcal{R}$), $a_1(\cdot)$ and $a_2(\cdot)$ is a feed-forward network with output as the weight, and $\sum_{j=1}^{K}w_{1j} =1$,$\sum_{j=K}^{2K}w_{2j} =1$. 

Leveraging the interest representation $\mathcal{R}(i_c)$, along with the $\textit{User Profile}\ X^U$, the $\textit{Item profile}\ X^I$ and $\textit{Context features}\ X^S$, we train the MFN model by optimizing the cross entropy loss
\begin{equation}
    \mathcal{L} = -\frac{1}{N}\sum_{x,y\in \mathbb{D}} (y\log p(x)+(1-y)\log(1-p(x))),
\end{equation}
where $\mathbb{D}$ is the training set, $x$ is represented by $[X^U, X^I, X^S]$ that is the input of the network, $y\in \{0,1\}$ denotes the labels, where $1$ means the user has an inveraction with the candidate item, $p(\cdot)$ is the final output of the model which represents the output probability.

\section{experiments}
In this section, we conduct experiments on 
both public and industrial datasets to evaluate the effectiveness of the proposed MFN compared with other representative methods. We  introduce the datasets, competitors, evaluation metrics and then report and analyse the experimental results.

\subsection{Experimental setup}
In this paper, we evaluate the model on both the public and industrial datasets. We collect the industrial datasets from an e-commerce app and leverage 3 days' logs with users' historical sequences for constructing the
training dataset and use the next one day's data for testing. For evaluating the performance on public datasets, the Area Under the Curve (AUC) is exploited and for the industrial datasets we use the RelaImpr \cite{DBLP:conf/icml/YanLXH14} to measure the relative improvement, which is defined as 
\begin{equation}
    RelaImpr = \left(  \frac{AUC(Test\_model)-0.5}{AUC(Base\_model)-0.5} -1    \right)\times 100 \%.
\end{equation}

We conduct our experiments on Tensorflow 1.4 and use the Adam optimizer. The initial learning rate is 1e-4, the epoch is set to 1. Other methods are set as their recommend settings.

\subsection{Experimental results}
The results on the industrial datasets is illustrated as Table. \ref{tab:1}

\begin{table}
  \caption{Experimental results on industrial datasets}
  \label{tab:1}
  \begin{tabular}{cl}
    \toprule
    Methods&RelaImpr\\
    \midrule
    Base    & 0.00\%  \\
    MIND     &   1.37\% \\
    ComiRec  &    1.39\% \\
    TransformerRec  & 1.47\% \\
    MFN w/o pretrain  & 1.50\% \\
    MFN w/o combination & 1.71\%\\
    MFN &     2.20\% \\
  \bottomrule
\end{tabular}
\end{table}

\section{Conclusion}
In the paper, we present Multiple interest and Fine granularity Network (MFN), which tackle users’ multiple and fine-grained interests and construct the model from both the similarity relationship and the combination relationship among the users’ multiple interests. We evaluate MFN on both public and industrial datasets. The experimental results demonstrate that the proposed MFN achieves superior performance than other existed representing methods.
\bibliographystyle{ACM-Reference-Format}
\bibliography{sample-base}










\end{document}